\documentclass[aps,twocolumn,secnumarabic,superscriptaddress,balancelastpage,amsmath,amssymb,nofootinbib,titlepage]{revtex4-2}
\usepackage[english]{babel}
\usepackage[utf8]{inputenc}
\usepackage{amsmath}\usepackage{bbold}
\usepackage{graphicx}
\usepackage[colorinlistoftodos]{todonotes}
\usepackage{bibentry}
\usepackage{url}
\usepackage{rotating}
\usepackage{float}
\usepackage{color}
\usepackage[sort&compress]{natbib}
\usepackage[colorlinks=true,citecolor=blue]{hyperref}
\usepackage[caption=false]{subfig}
\usepackage{empheq}
\usepackage{bm}
\usepackage[normalem]{ulem}
\usepackage{siunitx}
\usepackage{comment}

\newcommand{\bS}{{\bf S}}
\newcommand{\bJ}{{\bf J}}

\newcommand{\Uh}{\hat{U}}
\newcommand{\ab}{a_{\rm B}}
\newcommand{\Xh}{\hat{X}}
\newcommand{\Th}{\hat{T}}
\newcommand{\Ph}{\hat{P}}
\newcommand{\bh}{\hat{b}}
\newcommand{\ah}{\hat{a}}
\newcommand{\as}{a_s}
\newcommand{\Ah}{\hat{A}}
\newcommand{\Bh}{\hat{B}}
\newcommand{\bone}{\mathbb{1}}

\newcommand{\mb}{\bar{m}}

\newcommand{\bsigma}{\bm{\sigma}}

\newcommand{\nb}{n_{\rm B}}
\newcommand{\be}{\begin{equation}}
\newcommand{\ee}{\end{equation}}
\newcommand{\bea}{\begin{eqnarray}}
\newcommand{\eea}{\end{eqnarray}}
\newcommand{\bse}{\begin{subequations}}
\newcommand{\ese}{\end{subequations}}

\begin{document}
\title{Phonon Dynamics in Spherically-Curved Analog-Gravity Bose-Einstein Condensates}

\author{J. Austin Chunn}
\affiliation{Department of Engineering and Physics, Harding University, Searcy, AR, 72149 USA}
\affiliation{Department of Physics and Astronomy, Louisiana State University, Baton Rouge, LA 70803 USA}

\author{Ruotong Zhai}
\affiliation{Department of Physics and Astronomy, Louisiana State University, Baton Rouge, LA 70803 USA}

\author{Daniel E. Sheehy}
\affiliation{Department of Physics and Astronomy, Louisiana State University, Baton Rouge, LA 70803 USA}

\date{August 5, 2025}

\begin{abstract}
 We study the low energy phonon dynamics of a Bose-Einstein condensate (BEC) with a density profile that is equivalent, via a coordinate transformation, to phonons traveling in  a \lq\lq spherical\rq\rq\ curved spacetime that realizes the Friedman-Lema\^itre-Robertson-Walker (FLRW)
 metric.  The metric of this BEC is characterized by its curvature $\kappa$ and a time-depdendent scale factor $a(t)$, with an increase in the latter
 corresponding to an expansion of the analog FLRW universe.  We study 
  the propagation of classical phonons in such BECs, finding that a sudden change in the scale factor induces ripples in the wave motion.  In addition,
  we study  quantum phonon creation (or vacuum amplification) due to the
  scale-factor modification and quantify their entanglement.

\end{abstract}

\maketitle

\section{Introduction}
\label{sec:intro}

The field of analog gravity~\cite{Barcelo:2005fc} has attracted much recent
interest as a way to study phenomena from the realm of astrophysics
in a laboratory setting, such as a cold-atom or condensed matter 
experiment.  In addition to exploring phenomena that are normally not easily
accessible (such as black holes or the expansion of the early universe),
this field can strengthen ties between disparate fields of physics, including
cosmology, condensed-matter physics, cold atom physics, and quantum
information theory.

Some of the the recent setups that have been explored theoretically
and experimentally include the study of analog curved spacetimes for
Bose-Einstein condensates (BECs)~\cite{JainPRA}, 
superconducting circuits~\cite{Nation:2011dka},
the
dynamical Casimir effect~\cite{Jaskula2012},
Sakharov oscillations ~\cite{Hung:2012nc},
rotational superradiance ~\cite{Delhom},
as well as black hole physics and Hawking radiation~\cite{Philbin:2007ji,Belgiorno:2010wn,Weinfurtner:2010nu,Steinhauer:2015saa,Falque}.  The related Unruh effect
has been studied in
 a variety of settings,  such as  $^3$He~\cite{Volovik1992}, cold atomic gases 
\cite{Fedichev2003,Fedichev2004,Retzker:2008,Boada:2010sh,Rodriguez-Laguna:2016kri,Kosior:2018vgx,Hu:2018psq,Gooding:2020scc},
electrons in strong laser fields~\cite{Schutzhold2006}, 
hadron production in high-energy physics~\cite{Castorina2007,Castorina2022}, graphene~\cite{Iorio:2011yz,Iorio:2013ifa,Cvetic:2012vg,Bhardwaj2023,Tallent2024}, 
Weyl semimetals~\cite{Volovik:2016kid}, and 
quantum hall systems~\cite{Hegde:2018xub,Subramanyan:2020fmx}.
We also note that the expansion of the universe and
the theory of inflation in the context of analog systems has 
attracted much interest~\cite{Fischer:2004bf,Prain:2010zq, Eckel:2017uqx,Wittemer2019,Banik:2021xjn,
Llorente:2019rbs,
Bhardwaj:2020ndh,Bhardwaj2024}.  

The present work is inspired by a recent experiment by Viermann et al~\cite{Viermann}, with accompanying theoretical work in Refs.~\onlinecite{TS,SK2022}. These authors implemented 2D spherical and hyperbolic geometries, described by a 
Friedman-Lema\^itre-Robertson-Walker (FLRW)
metric,
using harmonically trapped  BECs with tailored density distributions. By applying power-law ramps to the scale factor of the BECs and comparing the results with analytical predictions, they observed signatures of pair particle production and the propagation of Sakharov oscillations. The strong agreement between experimental data and theory demonstrates that such configurable BEC systems serve as effective quantum simulators.

The aim of our work is to extend the theoretical framework to explore additional features of the 2+1D FLRW quantum field simulator. In particular, we investigate wave propagation in a 2D spherical spacetime with a time-dependent scale factor.  To do this, we compute the Greens function for the 
wave equation in the analog expanding curved spacetime BEC.  
As a fundamental tool for describing classical wave propagation, the Green’s function allows us to determine the wave response to any source via convolution.
One particular feature we find in such classical wave propagation is that 
additional ripples are induced by the onset (or end) of the expansion
(characterized by the scale factor) of the analog BEC.  
In addition, we study quantum particle production 
(or vacuum amplification)
in this analog
system, induced by modification of the scale factor entering
the FLRW metric, and quantify the entanglement of these particles.

The remainder of this paper is organized as follows: In Sec.~\ref{sec:BECANALOG}, we review the foundational setup of
the experiment by Viermann et al~\cite{Viermann}, including the realization of a 2D spherical FLRW universe via spatial modulation of the local density in a 2D BEC, and the derivation of the Klein-Gordon equation through the acoustic metric. 
Section~\ref{sec:waveprop} presents the calculation of the Greens function for the 2D wave equation in spherical spacetime. 
In Sec.~\ref{sec:StaticCase} we study waves in
a static curved-spacetime BEC, i.e., 
with a constant scale factor.  
In Sec.~\ref{sec:expanding}, we extend this analysis to cases with a time-dependent scale factor that simulates expansion of the
analog universe. We find that sudden changes in 
the scale factor time-dependence (at
the start and end of the analog expansion) 
lead to the formation of backward-
propagating ripples.
In Sec.~\ref{sec:partprod}, we study
quantum particle production due to
the change in scale factor in such BECs 
and quantify the particle entanglement.  
In Sec.~\ref{sec:concl} we provide some concluding
remarks.

\section{BEC and Analog Gravity}
\label{sec:BECANALOG}

 In this section, we review the setup of Viermann et al. who realized 
 analog 
 spherical and hyperbolic curved spacetimes in 
 a cold bosonic atomic gas~\cite{Viermann}.  Below we focus
 on the spherical case.  
We start with the 
following Hamiltonian for a 2D trapped interacting boson gas~\cite{PethickSmith}:
%
\bea
\hat{H}&=&\int d^2r\left(\hat{\Phi}^\dagger\hat{h}\hat{\Phi}+\frac{1}{2}U(t)\hat{\Phi}^\dagger\hat{\Phi}^\dagger\hat{\Phi}\hat{\Phi}\right),\\
\hat{h}&=&-\frac{\hbar^2}{2m}\nabla^2+V(r,t)-\mu.
\eea
Here  $\hat{\Phi}$ is the boson field operator and $\hat{h}$ is the single particle Hamiltonian with  chemical potential $\mu$ and with $\hbar$ being the reduced Planck's constant $m$ the atom mass. The term proportional to $U(t)$ describes the time-dependent interaction of the condensate atoms 
controllable by a Feshbach resonance.
The field operator dynamics follow from the Heisenberg equation of motion, which,
after writing $\hat{\Phi}=\phi_0+\delta\hat{\Phi}$ with $\delta\hat{\Phi}$ representing
fluctuations around the mean $\phi_0$, takes the form (to the zeroth order
in small $\delta \hat{\Phi}$):
\begin{equation}
\label{eq2}
\begin{gathered}
i\hbar\partial_t\phi_0=\left(-\frac{\hbar^2}{2m}\nabla^2+V(r,t)-\mu +U(t)n_0(r)\right)\phi_0.
\end{gathered}
\end{equation}
This is the well-known Gross-Pitaevskii equation that describes the ground state of the BEC, where $n_0=|\phi_0|^2$ is the mean density of the condensate atoms.
Taking the left side to vanish and neglecting the Laplacian term on the right side
yields the Thomas-Fermi (TF) approximation result for the local density:~\cite{BaymPethick}
\begin{equation}
\label{Eq:makinguse}
n_0(r)=\frac{\mu-V(r,t)}{U(t)}.
\end{equation}
Next, following Refs.~\cite{Viermann,TS}, we assume that the single-particle potential
is of the form $V(r,t) = \frac{1}{2}m\omega^2(t) f(r)$,
characterized by a 
controllable time-dependent frequency parameter $\omega(t)$ and 
with spatial dependence
$f(r) = - 2r^2 - \frac{r^4}{R^2}$.
Then, introducing the central density $\bar{n}_0 = \frac{\mu}{U(t)}$ and taking the frequency
dependence to satisfy:
\be
\frac{\frac{1}{2}m\omega^2(t)}{U(t)}= \frac{\bar{n}_0}{R^2},
\ee
we get the inhomogeneous TF density profile
\begin{equation}
n_0(r)=\bar{n}_0\left(1-\frac{f(r)}{R^2}\right) = \bar{n}_0\big(1+\frac{r^2}{R^2}\big)^2,
\end{equation}
with the last step holding when we plug in the above choice for $f(r)$.

Having obtained the mean local density, the next step is to consider small fluctuations about
this mean, which describe low-energy excitations.  This is done by keeping
terms linear in $\delta\hat{\Phi}$ in the equation of motion, leading to:
\bea
&&i\hbar\partial_t\delta\hat{\Phi}=\left(-\frac{\hbar^2}{2m}\nabla^2+V-\mu\right)\delta\hat{\Phi}
\\
&&\qquad +U(t) n_0(r)\left(2\delta\hat{\Phi}+\delta\hat{\Phi}^{\dagger} \right).
\nonumber 
\eea
Subtracting or adding the previous equation from its adjoint,
and making use of Eq.~(\ref{Eq:makinguse}), 
gives two coupled equations for $\delta\hat{\Phi}$ and $\delta\hat{\Phi}^\dagger$:
\bse
\label{Eq:EOM}
\bea
&&i\hbar\partial_t\left(\delta\hat{\Phi}^{\dagger}+\delta\hat{\Phi}\right)=-\frac{\hbar^2}{2m}\nabla^2\left(\delta\hat{\Phi}-\delta\hat{\Phi}^{\dagger}\right),
\\
\nonumber 
&&i\hbar\partial_t\left(\delta\hat{\Phi}-\delta\hat{\Phi}^{\dagger}\right)=-\frac{\hbar^2}{2m}\nabla^2\left(\delta\hat{\Phi}^{\dagger}+\delta\hat{\Phi}\right) 
\\
&&\qquad +2U(t) n_0(r)\left(\delta\hat{\Phi}^{\dagger}+\delta\hat{\Phi}\right).
\eea
\ese
The complex fields can be redefined in terms of two real fields $\hat{\phi}$ and
$\hat{\phi}_1$ using the definitions 
$\delta\hat{\Phi}+\delta\hat{\Phi}^{\dagger}=\sqrt{2}\hat{\phi}_1$ and $\delta\hat{\Phi}-\delta\hat{\Phi}^{\dagger}=2\sqrt{m}\hat{\phi}$. 
Upon rewriting Eqs.~(\ref{Eq:EOM}) in terms of  $\hat{\phi}_1$ and $\hat{\phi}$, and neglecting the Laplacian acting on $\hat{\phi}_1$, 
(corresponding to a long-wavelength or low-momentum approximation), we get:
\bea
\label{eqn6}
i\partial_t\hat{\phi}_1&=&-\frac{\hbar}{\sqrt{2m}}\nabla^2\hat{\phi},
\\
\label{eqn7}
\hat{\phi}_1&=& \frac{i\hbar\sqrt{m}}{\sqrt{2}U(t)n_0(r)}\partial_t\hat{\phi}.
\eea
Taking the time derivative of Eq.~(\ref{eqn7}) 
and combining with Eq.~(\ref{eqn6}) gives a Klein-Gordon equation of motion for the field $\hat{\phi}$:
\begin{equation}
-\partial_t\frac{1}{c_s^2(r)}\partial_t\hat{\phi}+\nabla^2\hat{\phi}=0,
\label{Eq:EOMKG}
\end{equation}
where $c_s(r)=\sqrt{\frac{U(t)n_0(r)}{m}}$ is the local wave speed.

\subsection{FLRW metric in spherical spacetime}
The next step is to show how the preceding system geometry,
leading to Eq.~(\ref{Eq:EOMKG}),
leads to an emergent 
FLRW metric for a expanding universe with 
spherical curvature.  
To do this, we rewrite the equation of motion in  tensor notation where the indices range from 0-2:
\begin{equation}
\label{curvewave}
\partial_\mu\left( \sqrt{|g|}g^{\mu\nu}\partial_\nu\hat{\phi}\right)=0.
\end{equation}
This equation is equivalent to the equation for a massless scalar field in a curved spacetime determined by the metric tensor $g_{\mu\nu}$ which is a three-by-three matrix. The components of the tensor are $\sqrt{|g|}g^{tt}=-1/c_s^2$ and $\sqrt{|g|}g^{ij}=\delta_{ij}$, where
$|g|=1/c_s^4$ is the absolute value of the determinant of the metric tensor. 

Now that the metric tensor has been determined, it is possible to find the metric line element for the analog space. The coordinates are the standard polar coordinates of radius and angle $(r, \varphi)$ paired with a time coordinate.   We get:
\begin{equation}
ds^2=g_{\mu\nu}dx^\mu dx^\nu=-dt^2+\frac{1}{c_s^2}\left(dr^2+r^2d\varphi^2\right),
\end{equation}
or equivalently, 
\begin{equation}
ds^2=-dt^2+a^2(t)\left(1-\frac{f(r)}{R^2}\right)^{-1}\left(dr^2+r^2d\varphi^2\right),
\end{equation}
where in the latter expression we have introduced $a^2(t)=\frac{m}{\bar{n}_0U(t)}$, the scale factor, anticipating
the subsequent mapping the to the  FLRW metric 
(where the scale factor describes the expansion of space).  
To arrive at this last step, we plug in the abovementioned choice for
$f(r) = -2r^2-\frac{r^4}{R^2}$ and make the variable change $u=\frac{r}{1+ \frac{r^2}{R^2}}$,
which leads to the 2D FLRW metric:
\begin{equation}
ds^2=-dt^2+a^2(t)\left(\frac{du^2}{1-\kappa u^2}+u^2d\varphi^2\right),
\end{equation}
The parameter $\kappa=4/R^2$ defines the curvature of the spacetime with positive curvature realizing spherical spacetime.

At this point we have established the correspondence between the 
experimental coordinate \(r\) and the radius for the acoustic metric \(u\). In order to solve the Klein-Gordon equation Eq.~(\ref{curvewave}), we make one more variable change from $u$ to an angle $\theta$:
\begin{equation}
\label{sphericalmap}
u= \frac{\sin\theta}{\sqrt{\kappa}}.
\end{equation}
With this variable change, the wave equation in 
the presence a scale factor \(a(t)\)
takes
the form 
\begin{equation}
\label{Eq:KGnew}
\partial_t\left(a^2(t)\partial_t\hat{\phi}\right)-\Delta\hat{\phi}=0,
\end{equation}
with $\Delta$ the Laplace-Beltrami operator in positively curved spacetime:
\begin{equation}
\Delta =\kappa \left[\frac{1}{\sin\theta}\partial_\theta\left(\sin\theta\,\partial_\theta\right)+\frac{1}{\sin^2\theta}\partial_\varphi^2\right].
\label{Eq:sphericalLap}
\end{equation}
In positive spacetime curvature, the eigenfunctions for the Laplace-Beltrami operator are modified spherical harmonic functions~\cite{TS}:
\begin{equation}
\Delta Y_{\ell m}(\theta, \varphi)=h(k)Y_{\ell m}(\theta, \varphi).
\label{eq:whyellm}
\end{equation}
Here $h(k)=-k(k+\sqrt{\kappa})$ is the eigenvalue of the operator and $k=\sqrt{\kappa}\ell$ is the wavenumber of the modes with $\ell$ 
an integer from $0$ to $\infty$. The spherical harmonics are related to the Legendre polynomials $P_{\ell m}(x)$ by the relationship.
\begin{equation}
\label{sph}
Y_{\ell m}(\theta, \varphi)=\sqrt{\frac{(\ell-m)!}{(\ell+m)!}}e^{im\varphi}P_{\ell m}(\cos\theta).
\end{equation}
Since the spherical harmonics form a complete basis over the analog space, it is possible to represent functions of the angular coordinates as linear combinations of these functions.

To simplify our numerical
calculations of wave propagation, below  
$\kappa$ is typically set to unity which means that $R=2$, this means that 
length scales for our calculations are given in units of $R/2$ or $\frac{1}{\sqrt{\kappa}}$. The scale factor $a(t)$ is defined to have units of $s/m$, inverse velocity.  Below, we use the initial
scale factor $a_i$ to define a characteristic timescale 
 $\frac{Ra_i}{2}$ or $\frac{a_i}{\sqrt{\kappa}}$, and measure
 time relative to this scale in our numerical calculations of classical
 wave propagation in the BEC.

\section{Wave propagation}
\label{sec:waveprop}
Our next task is to 
analyze {\em classical} wave propagation 
in the analog spherical curved spacetime, i.e., sound propagation in the 
case of the Laplacian operator given in Eq.~(\ref{Eq:sphericalLap}) and
in the presence of a time-dependent scale factor $a(t)$.  %
The classical limit corresponds to replacing the operator $\hat{\phi}$
in Eq.~(\ref{Eq:EOMKG}) by its expectation value.  The purpose
of this section is to set up the formalism, with the cases of 
static and expanding spacetimes studied in the subsequent sections Sec.~\ref{sec:StaticCase} and \ref{sec:expanding}.

\begin{figure}[!htbp]
  \centering
  \includegraphics[width=\columnwidth]{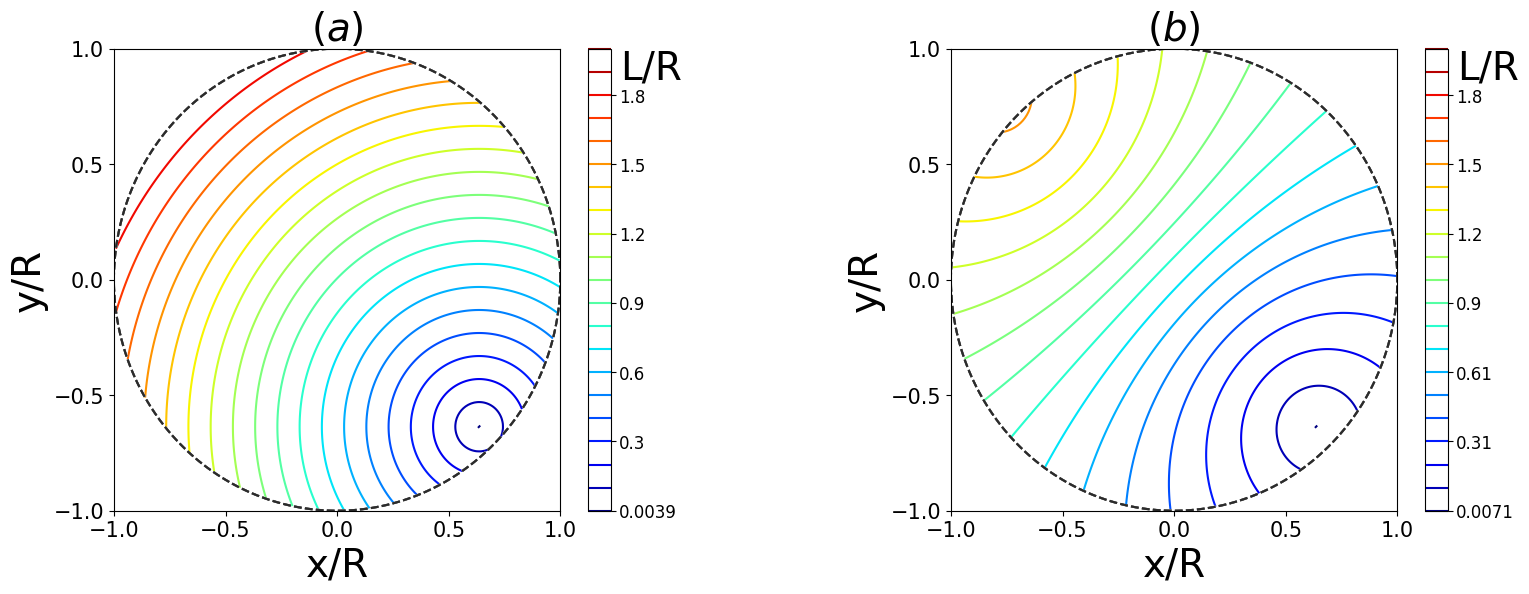}
  \caption{The figure depicts the equidistant lines
 in the experimental coordinates. Subplot (a) plots the equidistant lines in a flat geometry from the point $(r_0,\varphi_0)$ where $r_0=0.9R$ with $R$ being the total radius of the condensate and $\varphi_0=\frac{7}{4}\pi$. Subplot (b) plots the equidistant lines from the same point along the surface of a sphere projected onto a disc using Eq. ~ (\ref{sphericalmap}).}
  \label{comoving}
\end{figure}

The key 
quantity of interest is the wave equation Greens function
$G(\theta,\varphi,\theta_0,\varphi_0,t)$, which satisfies 
\begin{equation}
\label{eq:greenEOM}
\partial_t\left(a^2(t)\partial_tG\right)-\Delta G
= \frac{1}{\sin\theta}\delta(\theta-\theta_0)\delta(\varphi-\varphi_0)\delta(t),
\end{equation}
with the right side implementing a delta-function initial condition at time $t=0$.
In terms of $G$, the wave solution in the case of given initial conditions 
is~\cite{FollandPDE} 
(defining the abbreviated notation $\int_{\theta',\varphi'} \equiv \int_0^{2\pi} d\varphi' \int_0^{\pi/2}
d\theta'\sin\theta' $):
\bea
&&u(\theta,\varphi,t) = \partial_t \int_{\theta',\varphi'}
G(\theta,\varphi,\theta',\varphi',t)
u_0(\theta',\varphi')\nonumber 
\\
&&
+ \int_{\theta',\varphi'} G(\theta,\varphi,\theta',\varphi',t)
v_0(\theta',\varphi') ,
\label{Eq:22}
\eea
with the initial conditions
\bse
\label{eq:initialconditions}
\bea
&&u(\theta,\varphi,0) = u_0(\theta,\varphi),
\\
&&\partial_t u(\theta,\varphi,t)|_{t\to 0} = v_0(\theta,\varphi).
\eea
\ese
Thus, physically we can think of $G$ as  describing the wave emerging
from the initial condition of $u_0(\theta,\varphi) =0$
and $v_0(\theta,\varphi)
=\frac{1}{\sin\theta} \delta(\theta - \theta_0)
\delta(\varphi-\varphi_0)$.  

To compute $G$, we use the fact that  the $Y_{\ell m}(\theta,\varphi)$ are complete functions with the completeness relation:
\begin{multline}
\frac{1}{\sin\theta}\delta(\theta-\theta_{0})\delta(\varphi-\varphi_{0}) \\= \sum_{\ell=0}^{\infty}\frac{\ell+\frac{1}{2}}{2\pi} \sum_{m=-\ell}^{\ell}Y_{\ell m}(\theta,\varphi)Y_{\ell m}^{*}(\theta_0, \varphi_0).
\end{multline}
Using this, we can similarly write an expression for
$G$ in terms of a sum over the $Y_{\ell m}(\theta,\varphi)$:
\begin{equation}
\label{G general}
G = \sum_{\ell=0}^{\infty}\frac{\ell+\frac{1}{2}}{2\pi} \sum_{m=-\ell}^{\ell}F_k(t)Y_{\ell m}(\theta,\varphi)Y_{\ell m}^{*}(\theta_0, \varphi_0).
\end{equation}
where $F_k(t)$ are time-dependent coefficients that satisfy:
\begin{equation}
\partial_t\left(a^2(t)\partial_t F_k(t)\right)+|h(k)|F_k(t)=\delta(t).
\label{Eq:forEff}
\end{equation}
Equation~(\ref{Eq:forEff}) can be solved by letting $F_k(t)=\Theta(t)p_k(t)$ where $p_k(t)$ satisfies the initial conditions $p_k(0)=0$ and $\partial_tp_k(0)=\frac{1}{a^2(0)}$ and $p_k(t)$ solve the following homogeneous differential equation:
\begin{equation}
\label{chareq}
\partial_t\left(a^2(t)\partial_t p_k(t)\right)+|h(k)|p_k(t)=0.
\end{equation}
Solving Eq.~(\ref{chareq}) will produce the time-dependent coefficients of the wave equation which can be used to compute the numerical form of $G$ for a given choice for the scale factor $a(t)$. 
Below, we will focus on two specific cases: In Sec.~\ref{sec:StaticCase}
we study a static scale factor, representing a non-expanding spherical universe.  And, in Sec.~\ref{sec:expanding}, we study an expanding
universe, in which  $a(t)$ is initially static (given by $a(t) = a_i$), followed by expansion (represented by an exponentially increasing
$a(t)$) and a subsequent final static region (in which $a(t) = a_f$).

Before proceeding to these cases, we first simplify
our expression a bit more by plugging in Eq.~(\ref{sph})  for
the spherical harmonics to further simplify the Green function.
We obtain the simplified expression:
%
%
\begin{equation}
\label{GasL}
G=\sum_{\ell}\frac{\ell+\frac{1}{2}}{2\pi}F_k(t)P_{\ell}
(\cos(L)),
\end{equation}
where $L(\theta,\varphi,\theta_0,\varphi_0)$ is~\cite{SK2022}:
\bea
\label{Eq:comoving}
&&L(\theta,\varphi,\theta_0,\varphi_0)
\\
&&=
\cos^{-1}\left(\cos\theta\cos\theta_0+\sin\theta\sin\theta_0
\cos\left(\varphi-\varphi_0\right)\right),
\nonumber 
\eea
the comoving distance between the points $(\theta,\varphi)$ and $(\theta_0,\varphi_0)$. This expression for \(L\) matches exactly the geodesic distance on a two-dimensional sphere of unit radius, between two points with angular coordinates \((\theta,\varphi)\) and \((\theta_0,\varphi_0)\). It is important to note that while \(\varphi\) corresponds to the physical azimuthal angle in the lab frame, the \(\theta\) coordinate does not represent a true angular position in the lab. Instead, it arises from the transformation given in Eq.~(\ref{sphericalmap}), which relates to the radial position on the disk. This precise correspondence between the comoving length and the geometry of a 2D sphere highlights that, under the transformation, the lab frame effectively behaves as an analogue of a spherical universe.

In Fig.~\ref{comoving}, we provide an illustration of the comoving distance as observed in the lab frame.  Panel a shows the flat-spacetime case, for
which curves of equal comoving distance form circles (as expected).  
In panel b, we show the case of a spherical curved spacetime, with 
curves of equal comoving distance being distorted relative to the flat case
due to the underling curvature.  In each case, we expect classical wavefronts
from a point source to lie on these curves.
Having found the Green function for wave motion, our next task is to study wave evolution predicted by Eq.~(\ref{GasL}) in static ($a(t)$ constant) and dynamic (changing $a(t)$) spacetimes, which we address in the following sections.  In our calculations, we typically cut off the sum in Eq.~(\ref{GasL}) at a large value $\ell_{max}$.

\section{Static Spacetime Wave Propagation}
\label{sec:StaticCase}
\begin{figure}[!htbp]
  \centering
  \includegraphics[width=0.9\columnwidth]{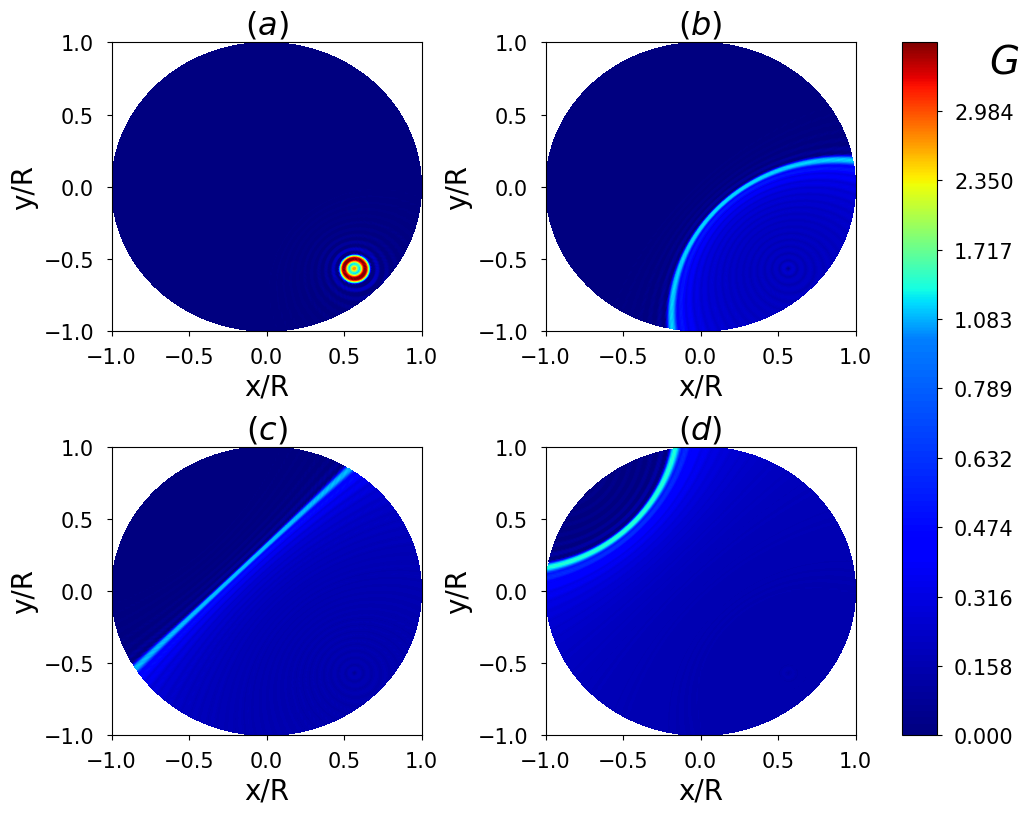}
  \caption{Green's function of the wave equation with static spacetime with $R=2$, $a_i=1$, and $\ell_{max}=100$.  The unit choices  for our dimensionful
  parameters are described at the end of Sec.~\ref{sec:BECANALOG}. Subplot (a) shows the wave at $t=0.1$, (b) shows $t=1$, (c) shows $t = 1.8$, and (d) shows $t= 2.5$.}
  \label{realwave}
\end{figure}
We start by studying wave propagation in the case of a static
scale factor (i.e., a non-expanding spherical analog universe).
For this case we take $a(t) = a_i$ and 
Eq.~(\ref{chareq}) is solved by:
\be
p_k(t)=A_k\sin\left(\omega_k^it\right)+B_k\cos\left(\omega_k^it\right),
\ee
where $\omega_k^i=\frac{\sqrt{|h(k)|}}{a_i}$ is the phonon mode frequency at wavenumber $k$.
The value  of the coefficients $A_k$ and $B_k$ can be determined by the initial conditions,
which is simply 
$p_k(0)=0$ and $\partial_tp_k(0)=\frac{1}{a_i^2}$. The coefficients that satisfy this are $A_k=\frac{1}{a_i\sqrt{|h(k)|}}$ and $B_k=0$ and the time-dependent coefficients needed for the Greens function 
take the following form:
\begin{equation}
\label{F function}
F_k(t)=\frac{\Theta(t)}{a_i\sqrt{|h(k)|}}\sin(\omega_k^it).
\end{equation}
Upon plugging this into Eq.~(\ref{GasL}), we arrive at 
\begin{equation}
\label{Eq:finalgreen}
G=\sum_{\ell}
\frac{\ell+\frac{1}{2}}{2\pi}\frac{\Theta(t)}{a_i\sqrt{|h(k)|}}\sin(\omega_k^it)
P_{\ell}(\cos(L)).
\end{equation}
Since all coordinate dependence enters via $L$, we see that
wavefronts will form along curves of equal co-moving length from 
the initial wave source.

This is
illustrated in Fig.~\ref{realwave}, which shows  the wave
for different time steps starting from a point at radius $r=0.8R$ and angle $\varphi=\frac{7\pi}{4}$ with a scale factor $a_i=1$. 
In Fig.~\ref{realwave} the darker blue parts show the area of the condensate that is still undisturbed, while the brighter areas show the wave peak with a trailing
edge. The first plot, panel a, shows the system shortly after the wave is generated, and the successive plots show the wave at later times.  By
comparing to Fig.~\ref{comoving}, we can see that the wavefronts are indeed
along curves of equal comoving distance.

\begin{figure}[!htbp]
  \centering
  \includegraphics[width=0.9\columnwidth]{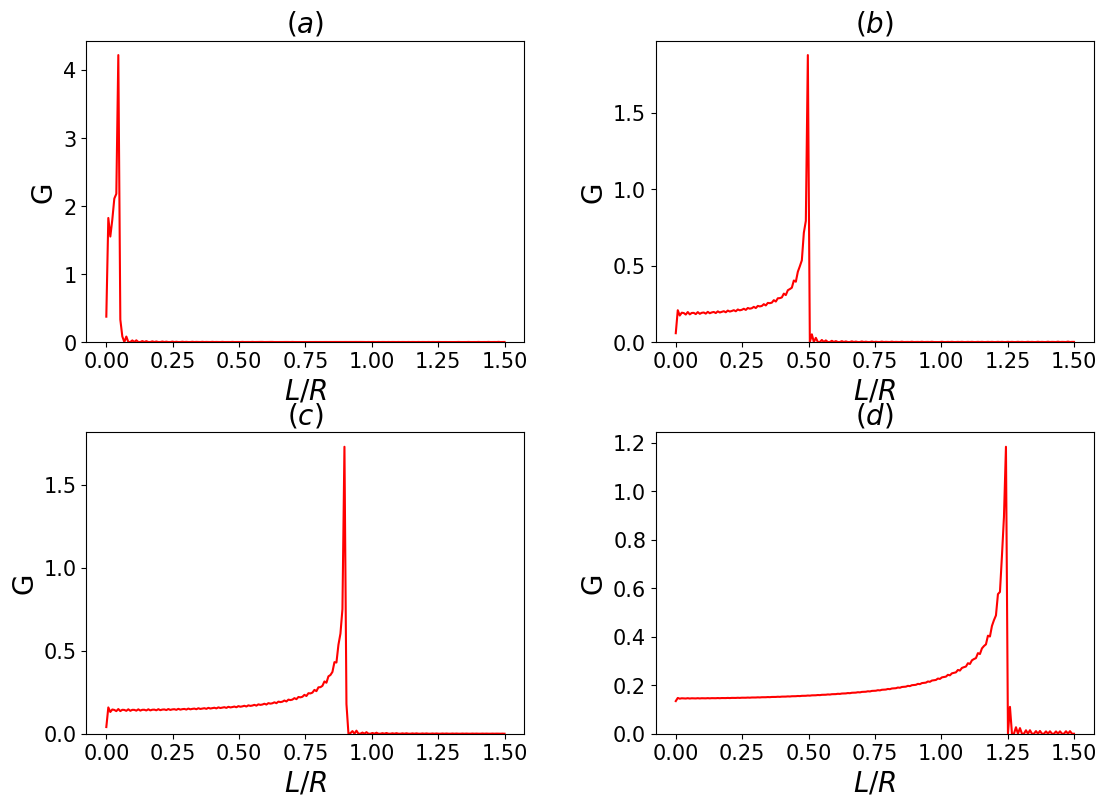}
  \caption{The panels depict the same wave as in 
  Fig.~\ref{realwave} showing the 
  shape vs. comoving distance $L$ at different time steps.  
  Here the parameters
    $R=2$, $\ell_{max}=1500$, and $a_i=1$. The subplot $(a)$ shows the wave at point $t=0.1$, $(b)$ shows the wave at $t=1$, $(c)$ shows the wave at $t=1.8$, and $(d)$ shows the wave at $t=2.5$.}
  \label{static as L}
\end{figure}

Since the spatial dependence of the wave is only via the comoving
distance, as seen in Eq.~(\ref{GasL}), henceforth we shall simply plot results for $G$ as a function 
of $L$, with the understanding that the full real-space form of the wave is
via Eq.~(\ref{Eq:comoving}).

Figure~\ref{static as L} plots the same parameters as Fig.~\ref{realwave} as a function of the comoving distance $L$ which shows how the amplitude varies in space.  We see the expected behavior in 2D of a sharp wavefront 
moving with time,  followed by
a trailing edge. 

\begin{figure}[h!]
  \centering
  \includegraphics[width=\columnwidth]{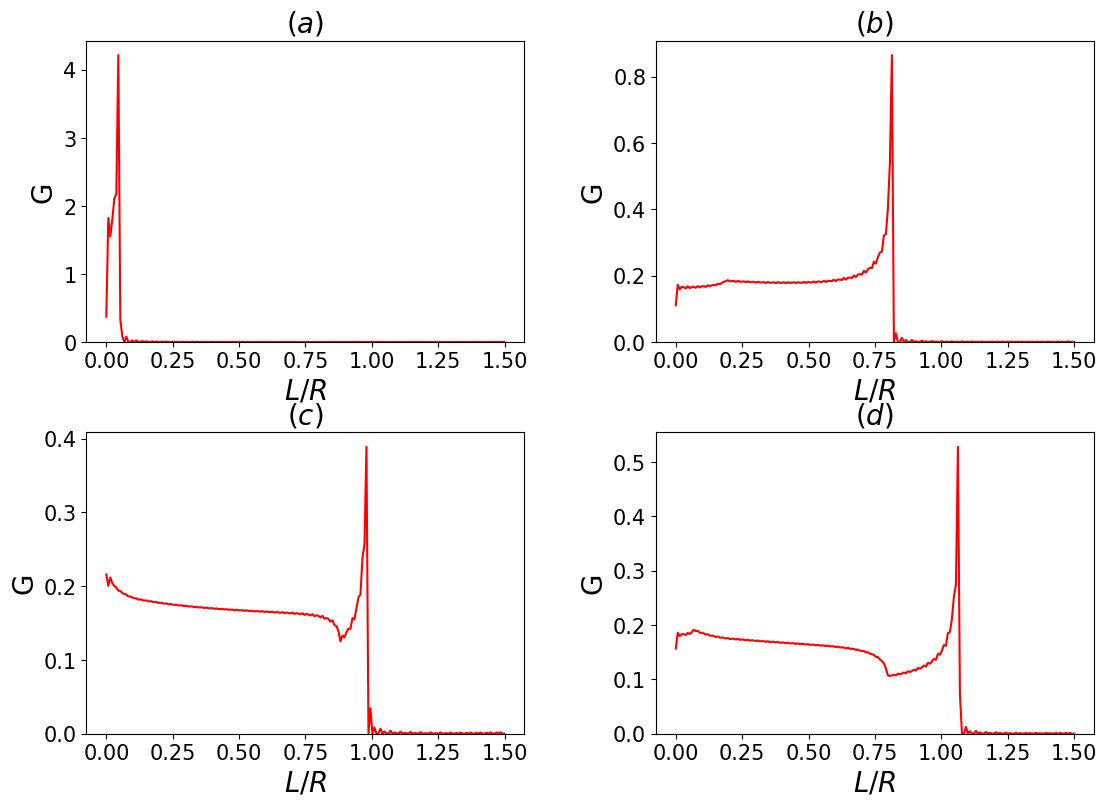}
  \caption{The figure depicts wave propagation in an analog expanding spherical BEC.  The panels show the
  wave amplitude as a function of $L$ for different
  time steps.  Here, the parameters are $H=1$, $t_i=1$, $t_f=3$, $\ell_{max}=1500$, $R=2$, and $a_i=1$. Subplot $(a)$ displays the wave at $t=0.1$  while $(b)$ shows the wave at $t=2$, $(c)$ shows the wave at $t=3.8$ and $(d)$ shows the wave at $t=5$.}
  \label{fig:sub1}
\end{figure}

\section{Expanding Spacetime}
\label{sec:expanding}
In this section, we consider an exponentially expanding spacetime (a de Sitter Universe). To model this, we consider
wave motion with the following time-dependent scale factor:
\begin{equation}
\label{Scale Factor}
a(t)= \begin{cases} 
      a_i & t\leq t_i, \\
      a_ie^{H(t-t_i)} & t_i\leq t\leq t_f, \\
      a_f & t\geq t_f, 
   \end{cases}
\end{equation}
where $a_f=a_ie^{H(t_f-t_i)}$ is the scale factor following expansion and $H$ defines the expansion rate and is analogous to the Hubble constant $H=\frac{\dot{a}}{a}$.

With this choice for the scale factor, the function $F_k(t)$ in Eq.~(\ref{GasL}) will also have three regimes, which we write as 
\begin{equation}
F_k(t)= \begin{cases} 
      \Theta(t)p_k^i(t) & t\leq t_i, \\
     p_k^{II}(t) & t_i\leq t\leq t_f, \\
      p_k^f(t) & t_f\leq t .
   \end{cases}
\end{equation}
In order to solve the wave equation for this expansion profile we need to impose the initial conditions $p_k(0)=0$ and $\partial_tp_k(0)=\frac{1}{a^2(0)}$. Prior to expansion the function $F_k(t)=\Theta(t)p^i_k(t)$ will take the form of equation \eqref{F function}. During the expansion,
for times $t_i\leq t\leq t_f$, the function 
$p^{II}_k(t)$ satisfies 
the following differential equation:
\begin{equation}
\partial_t\left(e^{2H(t-t_i)}\partial_t p^{II}_k(t)\right)+\left(\frac{\sqrt{|h(k)|}}{a_i}\right)^2p^{II}_k(t)=0,
\end{equation}
which has the following solution:
%
\bea
&&
\label{eq:peetoo}
p^{II}_k(t)=e^{-Ht}
\Big[A_iJ_1(\frac{\omega^I_k}{H}{\rm e}^{-H(t-t_i)})
\\
&&\qquad 
\nonumber 
+B_iY_1\big(\frac{\omega^I_k}{H}{\rm e}^{-H(t-t_i)})
\Big],
\eea
with $J_1$ and $Y_1$ Bessel functions.
The coefficients $A_i$ and $B_i$ are determined by matching
the solutions at time $t_i$, with our results for 
these found in Appendix~\ref{sec:appendixA}.

Finally, following
the expansion in the regime $t\geq t_f$,  the differential equation and solution are given by: 
\bea
&&\partial_t^2p^f_k(t)+\left(\frac{\sqrt{|h(k)|}}{a_f}\right)^2p^f_k(t)=0,
\\
&&p^f_k(t)=A_f\sin\left(\omega_k^ft\right)+B_f\cos\left(\omega_k^ft\right),
\label{Eq:peeEff}
\eea
where $\omega_k^f=\frac{\sqrt{|h(k)|}}{a_f}$ is the final frequency of the phonon modes (following expansion). The coefficients $A_f$ and $B_f$ can be again determined from 
matching the solution at time $t_f$, with their full form shown in 
Appendix~\ref{sec:appendixA}.

\begin{figure*}[!htbp]
  \centering
  \includegraphics[width=.8\linewidth]{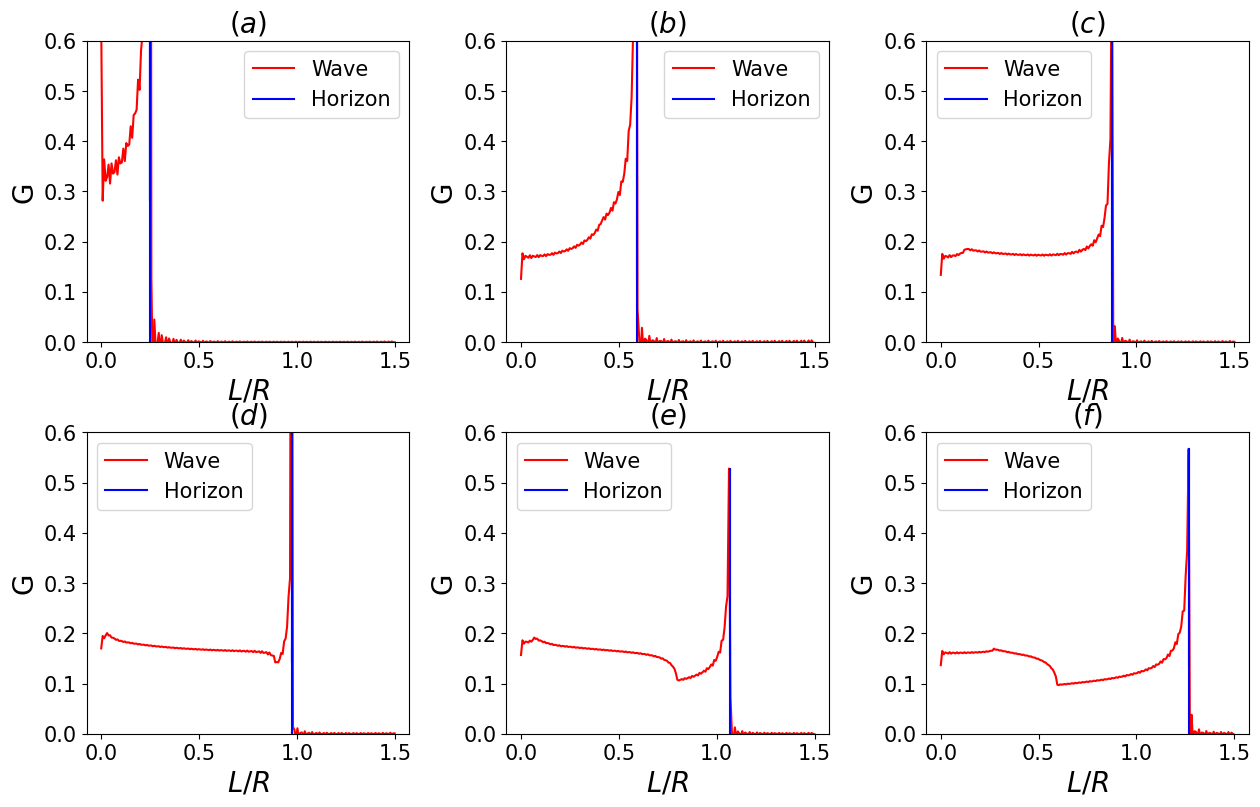}
  \caption{The six panels show the Greens function describing
  wave propagation, with the initial scale factor $a_i = 1$.  
  The increase in scale factor (with rate $H=1$) 
  starts at $t_i = 1$ and ends at $t_f = 3$. Subplot $(a)$ plots the wave at $t=0.5$, $(b)$ at $t=1.2$, $(c)$ at $t=2.4$, $(d)$ at $t=3.6$, $(e)$ at $t=5$, and $(f)$ at $t=8$.  In each 
  panel, the vertical blue line indicates the predicted location of the 
  wave edge according to Eq.~(\ref{Eq:resultforhvst}).   Note
 that the start of expansion creates a backward-traveling ripple that is
 positive relative to the background (seen in panels (b) and (c)), and the
 end of expansion creates a backward-traveling ripple that is seen in panels
 (d), (e), and (f).  Figures~\ref{fig:sub4} and \ref{fig:sub5} focus on the
 inital and final ripples, respectively.  
  }
  \label{figWide}
\end{figure*}

\subsection{Numerical Results}
Using the preceding analytic results for the function
$F_k(t)$, we can plug into Eq.~(\ref{GasL}) to
determine the time-dependent
Green's function during the expansion represented by
the scale factor $a(t)$ in Eq.~(\ref{Scale Factor}).
This calculation determines the profile and properties
of a classical wave created at time $t=0$, with a 
delta-function initial condition (See Eq.~\ref{eq:initialconditions}.)

As in the static spacetime case, it is sufficient to
consider the wave as a function of comoving distance
$L$.  
Figure~\ref{fig:sub1} depicts the shape of the wave 
as a function of comoving distance $L$ for the case of 
parameters 
 $H=1$, $R=2$, $t_i=1$, $t_f=3$, and $a_i=1$, with the sum 
in Eq.~(\ref{GasL}) evaluated numerically up to a 
cutoff $l_{max}=1500$.  Panel $a$ shows the wave
 prior to the expansion expansion regime (i.e. $t<t_i$), 
  panels $b$ and $c$ show the wave during expansion (i.e., for time $t$ satisfying $t_i<t<t_f$), and panel $d$ shows the wave after expansion $t>t_f$. 
  These panels show that wave has a sharp maximum value at a specific point $L_{m}$, such that for all distances greater than the peak $L>L_{m}$, the wave has zero amplitude, and for distances less than the peak $L<L_{m}$, the wave intensity has a reduced (but not vanishing)
  amplitude. In addition to these broad features,
  we observe a small additional ripple moving towards
  decreasing $L$ (i.e., propagating backwards) in panels
  b and c (i.e., during the expansion).  
  In addition, after expansion (seen in panel d) 
  we observe a negative-amplitude ripple (subtracting
  from the background tail of the wave) that also 
  propagates backwards.  We note that these ripples are so small
  that they might be mistaken for numerical errors due to the finite
  cutoff in our sums.  However, below we derive analytic formulas for
  the location of the main wave edge and of these ripples, confirming
  that the ripples are induced by the onset and end of the 
  inflation.

  To better understand the propagation of the main wave and ripples,
  in the next section we discuss particle horizons in
  the analog BEC.

\subsection{Particle Horizon}

Since  phonons in the BEC correspond to massless waves in the analog spacetime, they should propagate at the maximal speed of the space, which is given by the particle horizon, essentially following the local
\lq\lq speed of light\rq\rq\ or wave speed, which is generally given by 
$c/a(t)$ (note we have set the sound speed $c$ to unity).
According to Ellis and Rothman~\cite{Ellis1993}, the particle horizon for the FLRW metric is determined by the scale factor:
\begin{equation}
h(t)=\int_{0}^t\frac{dt'}{a(t')},
\end{equation}
which essentially integrates the local wave speed to find
the distance the wave propagates.  For the specific case considered
here, using Eq.~(\ref{Scale Factor}), we get for the horizon location:
\begin{equation}
\label{Eq:resultforhvst}
h(t)=
\begin{cases}
\frac{1}{a_i}t ,& t\leq t_i\\
\frac{1}{Ha_i}(1-e^{-H(t-t_i)})+\frac{1}{a_i}t_i ,& t_i\leq t\leq t_f\\
\frac{1}{a_i}e^{-H(t_f-t_i)}(t-t_f)\\+\frac{1}{Ha_i}(1-e^{-H(t_f-t_i)})+\frac{1}{a_i}t_i ,& t\geq t_f
\end{cases}
\end{equation}
which we claim should determine the wave edge in our calculations.

To demonstrate that the particle horizon formula in 
Eq.~(\ref{Eq:resultforhvst}) indeed 
determines the position of the wave edge, 
in Fig.~\ref{figWide} we show more results for the 
Green's function vs. $L$ for the case of the parameters $t_i = 1$, $t_f=3$, 
$H=1$, $a_i=1$, $R=2$, and $\ell_{max}=1500$,
with the vertical blue line showing the particle horizon 
$h(t)$ according to Eq.~(\ref{Eq:resultforhvst}). 
This shows that the edge of the wave matches exactly with the particle horizon, which tells us that the wave acts as a massless scalar field since it propagates at the same speed as the horizon. 
Figure~\ref{figWide} also shows the small backwards-propagating ripples
seen in Fig.~\ref{fig:sub1}.  To demonstrate that these features
are indeed waves, we can find horizon formulas for their 
position vs. time. 



To find formulas for the locations of the backwards traveling
ripples that occur in the time periods during and after 
expansion, we make one additional assumption.  We assume
that the origin of these ripples comes from the discontinuities
in the scale factor at times $t_i$ (when expansion starts) and
$t_f$ (when expansion ends).

\begin{figure}[!htbp]
  \centering
  \includegraphics[width=\columnwidth]{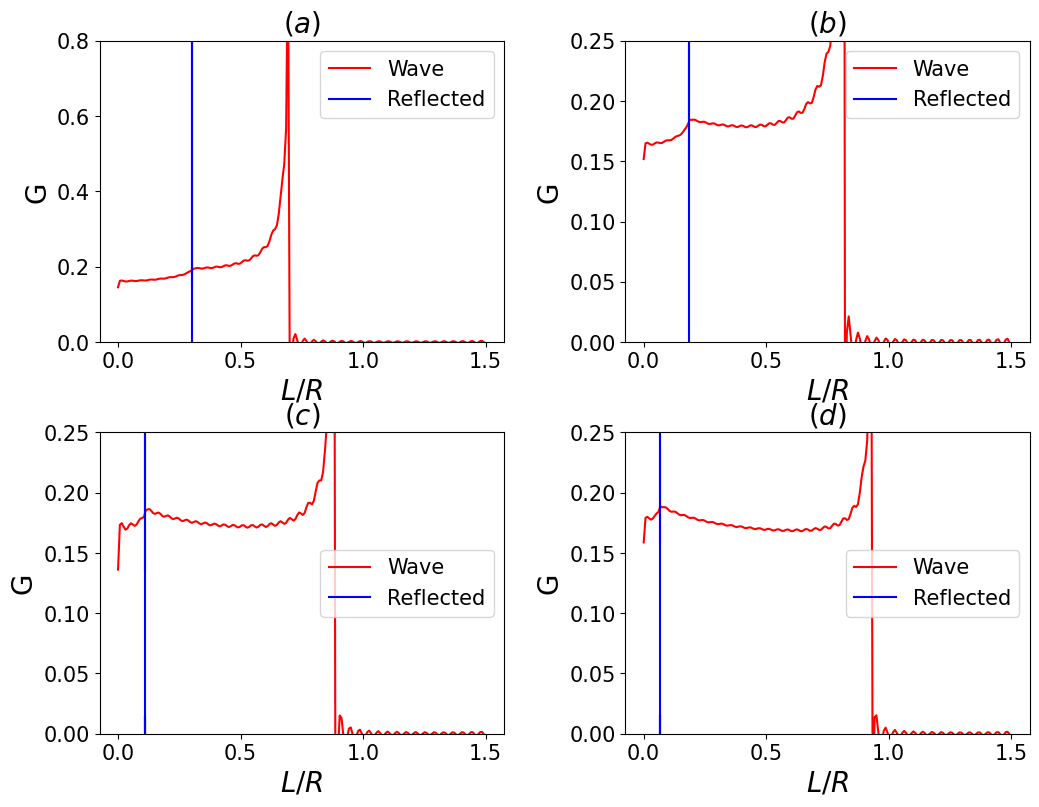}
  \caption{Particle Horizon of the Initial Reflected Wave (or ripple). Subplot $(a)$ shows the wave at $t=1.5$, $(b)$ at $t=2$, $(c)$ at $t=2.5$, and $(d)$ at $t=3$.
  The blue lines in each panel show the predicted location of
  a ripple sent backwards from the wave edge at time $t_i$ according to
  Eq.~(\ref{Eq:horizonripple1}).
  }
  \label{fig:sub4}
\end{figure}

If this assumption is valid, then the first backwards traveling
wave should be created at the location of the main wave edge
at time $t_i$ and propagate at the time-dependent speed of light $1/a(t)$.
This implies the particle horizon marking the location of the ripple is
given by:
\begin{equation}
\begin{gathered}
h_{r,1}(t)=-\frac{1}{Ha_i}\left(1-e^{-H(t-t_i)} \right)+\frac{1}{a_i}t_i.
\label{Eq:horizonripple1}
\end{gathered}
\end{equation}
As seen in Fig.~\ref{fig:sub4}, we indeed numerically find that the 
ripple location agrees with $h_{r,1}(t)$ (with the latter indicated by a vertical blue line.)
Note that, at $t= t_i$, $h_{r,1}(t)$ coincides with $h(t)$, i.e.,
as stated above, we
assume that the ripple emerges from the main wave edge at time $t_i$.  
We make a similar assumption for the ripple created at $t_f$ 
(i.e., at the end of the inflation.)  Via similar arguments, 
we find for the second ripple location: 
\begin{figure}[!htbp]
  \centering
  \includegraphics[width=\columnwidth]{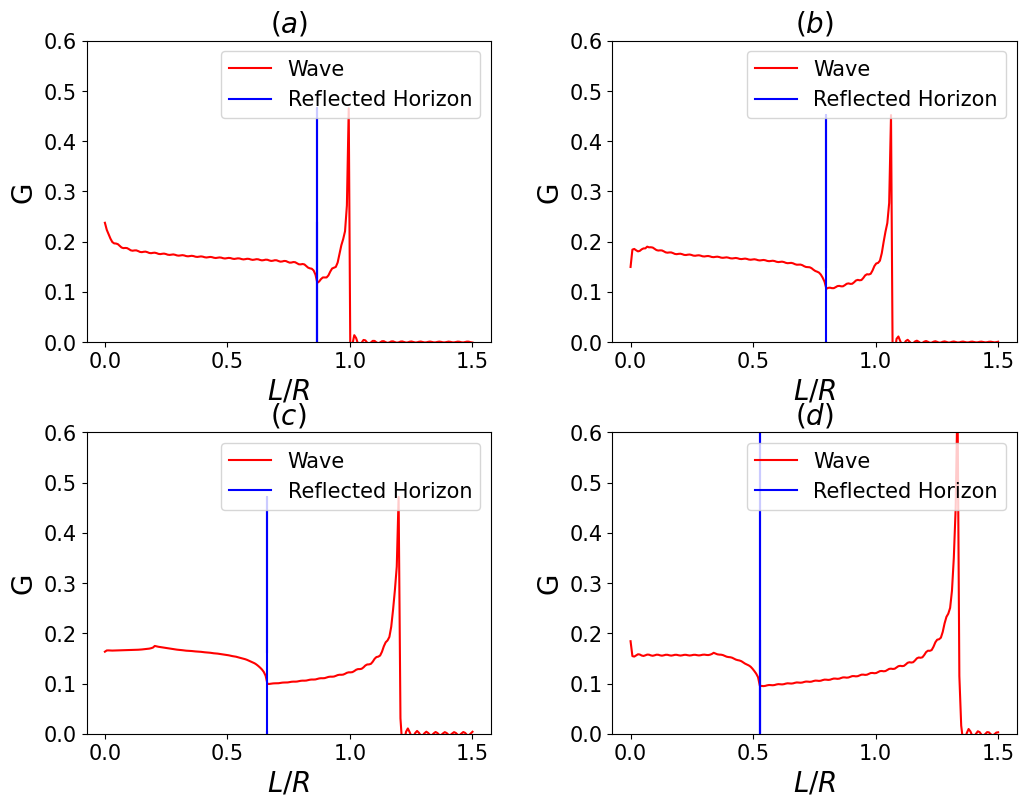}
  \caption{Particle Horizon of the Final Reflected Wave (or ripple). Subplot $(a)$ shows wave at $t=4$ seconds, $(b)$ at $t=5$ seconds, $(c)$ at $t=7$ seconds, and $(d)$ at $t=9$ seconds.
  The blue lines in each panel show the predicted location of
  a ripple sent backwards from the wave edge at time $t_i$ according to
  Eq.~(\ref{Eq:horizonripple2}).}
  \label{fig:sub5}
\end{figure}
\begin{equation}
\begin{gathered}
h_{r,2}(t)=-\frac{1}{a_i}e^{-H(t_f-t_i)}(t-t_f)\\
+\frac{1}{Ha_i}\left( 1-e^{-H(t_f-t_i)}\right)+\frac{1}{a_i}t_i.
\label{Eq:horizonripple2}
\end{gathered}
\end{equation}
In Fig.~\ref{fig:sub5}, we show the wave amplitude vs. $L$ 
for times after the end of the
expansion, i.e., $t>t_f$.  These panels show that, while the main wave still
propagates to the right, a negative ripple is emitted and propagates to the
left.  The blue lines indicate show the predicted location of this ripple
according to Eq.~(\ref{Eq:horizonripple2}), showing that (as predicted)
the observed ripple is not a numerical artifact since it propagates 
at the final speed of sound $c_f = 1/a_f^2$

These results show that the sound waves propagating in the BECs behave like massless waves in the expanding FLRW spacetime. The
agreement with the horizon formulas show that the ripples in the condensate are reflected waves and not artifacts of the calculation.
These results validate these BECs as simulators of spacetime expansion 
and indicates that they could be used for future tests of wave propagation
in curved spacetime.

\section{Quantum Particle Production}
\label{sec:partprod}
Having studied classical wave propagation in an analog trapped BEC that 
simulates a spherical expanding 2D universe, in the present section
we study the possibility of quantum particle (or phonon) production.
The setup we have in mind is as follows.  We assume an initially
prepared equilibrium
BEC, characterized by an initial sound speed and scale factor $a_i$.
This initial BEC will have a well-defined vacuum state annihilated by
corresponding mode operators.  Subsequently we assume a sudden 
analog expansion of the BEC, represented by a modification of the 
scale factor $a(t)$ according to Eq.~(\ref{Scale Factor}). 
The final
system at scale factor $a_f$ is characterized by a different final 
vacuum state 
with a different set of mode operators, with the dynamics
induced by $a(t)$ effectively implementing a two-mode squeezing operation
and concomitant particle production.  

To demonstrate this, we return to the Klein-Gordon 
Eq.~(\ref{Eq:KGnew}) for the BEC fluctuations and expand
the excitation field operator $\hat{\phi}$ as a sum over the eigenfunctions of $\Delta$:
\begin{equation}
\begin{gathered}
\hat{\phi} = \sum_{\ell=0}^{\infty}\frac{\ell+\frac{1}{2}}{2\pi}\left[\hat{a}_{km}Y_{lm}(\theta,\varphi)v_k(t)+\hat{a}^{\dagger}_{km}Y_{lm}^{*}(\theta,\varphi)v_k^{*}(t)\right].
\label{eq:operatorEquation}
\end{gathered}
\end{equation}
Here, the 
 operators $\hat{a}_{km}$ and $\hat{a}^{\dagger}_{km}$ 
 respectively annihilate and create phonon quanta, with the
 phonon vacuum defined by $\hat{a}_{km}|0\rangle =0$.

The mode functions $v_k(t)$ control the time dependence
of the field operators and are  determined by solving the mode equation:
\begin{equation}
\label{mode eq}
\ddot{v}_k(t)+2\frac{\dot{a}(t)}{a(t)}\dot{v}_k(t)+\frac{|h|}{a^2(t)}v_k(t)=0.
\end{equation}
Because of the canonical commutations of the operators, the mode functions must satisfy a normalization condition defined in terms of the Wronskian: 
\begin{equation}
\text{Wr}[v_k, v_k^{*}]=a^2(t)\left(v_k\dot{v}_k^{*}-\dot{v}_kv_k^{*}\right) = i .
\end{equation}
The scale factor given in equation \eqref{Scale Factor} defines three regions in which the mode functions take different forms. In regions I and III the scale factor is static, so the mode functions will obey a simplified equation:
\begin{equation}
\label{simp mode eq}
\partial_t^2v_k(t)+\frac{|h|}{a^2}v_k(t)=0.
\end{equation}
The solution to equation \eqref{simp mode eq} will
generally be a sum of complex exponential functions
frequency with $\omega_k=\frac{\sqrt{|h(k)|}}{a}$.  However,
in the initial regime where $a= a_i$, we know that
the vacuum state is annihilated by $\hat{a}_{km}$ and
quanta are created by $\hat{a}_{km}^\dagger$.  This implies
that the correct solution in region $I$ is:
%
%
\begin{equation}
\label{eq:vonequantum}
v^{\text{I}}_k(t)= \frac{{\rm e}^{-i\omega^i_kt}}{a_i\sqrt{2\omega^i_k}},
\end{equation}
with $\omega^i_k$ being $\omega_k$ with $a\to a_i$.
When we plug this into Eq.~(\ref{eq:operatorEquation}),
we see that this choice equivalently puts the mode operators
in the Heisenberg picture, i.e., 
$\hat{a}_{km} \to \hat{a}_{km}{\rm e}^{-i\omega^i_kt}$.

In Region II, the scale factor is time-dependent, and the mode functions obey a more complicated differential equation:
\begin{equation}
\begin{gathered}
\partial_t^2v_k(t)+2H\partial_tv_k(t)+|h(k)|e^{-2Ht}v_k(t)=0.
\end{gathered}
\end{equation}
The mode functions that solve this equation are a combination of Bessel functions of the first $J_n(x)$ and second $Y_n(x)$ kind:
\begin{equation}
\begin{gathered}
v_k^{II}(t)=e^{-Ht}\sqrt{|h(k)|}\Bigg[AJ_1\left(-\frac{\omega_k^i}{H}e^{-Ht}\right)\\
+iBY_1\left(-\frac{\omega_k^i}{H}e^{-Ht}\right)\Bigg],
\label{eq:gathered}
\end{gathered}
\end{equation}
with the coefficients $A$ and $B$ determined by enforcing
continuity of $v_k(t)$ and its derivative at the initial
time 
$v_k^{\text{I}}(t_i)=v_k^{II}(t_i)$ and $\dot{v}_k^{\text{I}}(t_i)=\dot{v}_k^{II}(t_i)$,
where we take $t_i =0$.  We find for the coefficients
$A$ and $B$:
\bea
A&=&\frac{(\omega_k^i)^{3/2} \pi \left( Y_0\left(\frac{\omega_k^i}{H}\right) - i Y_1\left(\frac{\omega_k^i}{H}\right) \right)}{2\sqrt{2} |h(k)| H}, \\
B&=& \frac{ \pi \left(i J_0\left(\frac{\omega_k^i}{H}\right) +  J_1\left(\frac{\omega_k^i}{H}\right) \right)}{2\sqrt{2}  Ha_0^2\sqrt{\omega_k^i}} .
\eea

In Region III the scale factor will be static again which means that the mode functions will obey equation \eqref{simp mode eq}.  However, to match solutions at $t_f$, 
the final $v_k^{III}(t)$ must generally be a sum of
complex exponential functions:
\begin{equation}
v_k^{III}(t) = \alpha_k^{*}
\frac{e^{-i\omega_k^ft}}{a_f\sqrt{2\omega_k^f}}
+\beta_k
\frac{e^{i\omega_k^ft}}{a_f\sqrt{2\omega_k^f}},
\label{eq:finalV}
\end{equation}
with the parameters $\alpha_k$ and $\beta_k$ satisfying
$|\alpha_k|^2 -|\beta_k|^2 = 1$.  The parameter
$\beta_k$ is the particle production amplitude, with
$|\beta_k|^2$ being the number of quanta excited with
wavevector $k$ during the expansion process.  To see
why this is the case, note that if we plug 
Eq.~(\ref{eq:finalV}) into Eq.~(\ref{eq:operatorEquation}),
the field operator in region III is the same as in
region I but with $\hat{a}_{km}$ replaced by new
(final system) mode operators $\hat{b}_{km}$ given by
\be
\label{eq:modoprel}
\hat{b}_{km} = \alpha^*_k \hat{a}_{km} + 
\beta^*_k \hat{a}_{k\mb}^\dagger,
\ee
a two-mode squeezing operation.  Here,
$\mb = -m$.
Working in the Heisenberg picture, if we assume an initial
vacuum state $|0\rangle$ (annihilated by the
$\hat{a}_{km}$) in region I, then the number of 
quanta at quantum numbers $k,m$ is 
\begin{equation}
N_{km}= \langle 0| \hat{b}^{\dagger}_{km}\hat{b}_{km}|0\rangle = |\beta_k|^2,
\end{equation}
giving the interpretation of $|\beta_k|^2$ as 
the number of created particles.  

\begin{figure}[!htbp]
  \centering
  \includegraphics[width=\columnwidth]{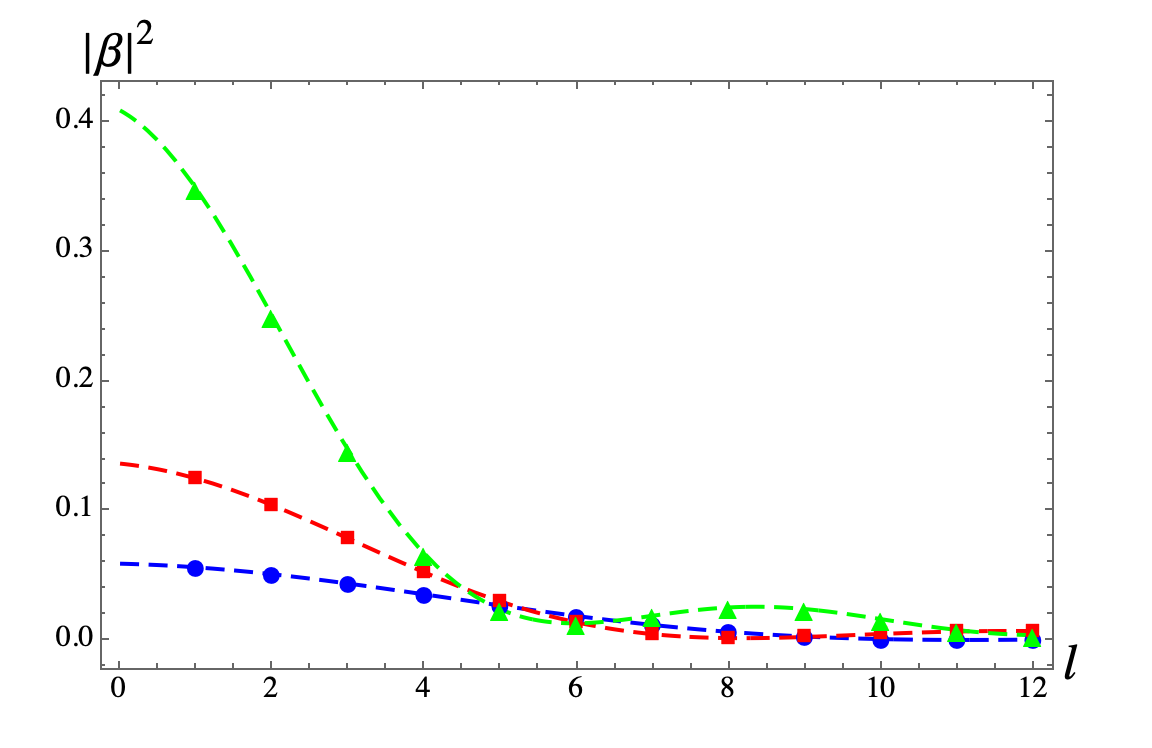}
  \caption{The particle production parameter \(|\beta|^2\) as function of mode index \(k\). For this plot, we took the initial scale factor \(a_0=331s/m\),
  Hubble rate \(H=240s^{-1}\), and various
  expansion time \(t_f=2ms\) (blue circles), \(t_f=3ms\) (red squares), and \(t_f=5ms\)(green triangles).   
  The longer \(t_f\) is, the larger the final scale factor is and more particles are produced.}
  \label{particleproduction}
\end{figure}

To compute the parameters $\alpha_k$ and $\beta_k$, 
we that the mode function and its derivative are
continuous at the end of expansion, i.e., 
$v^{II}_k(t_f)=v_k^{III}(t_f)$ and $\dot{v}_k^{II}(t_f)=\dot{v}_k^{III}(t_f)$.  For simplicity we leave the final result
for $\beta$ to Eq.~(\ref{eq:afourforbeta}) in 
Appendix~\ref{sec:appendixA}.  In Fig.~\ref{particleproduction} 
we show a plot of the particle-production number 
$|\beta|^2$ in for a few different
choices for the expansion time, with other parameter choices
described below.

To motivate the parameters used in our plot of
$|\beta|^2$, we 
turn to the experiment of Viermann et al~\cite{Viermann}
and try to obtain realistic estimates.  In this section we
restore dimensionful quantities previously set to unity
in order to estimate various parameters.  The first
such parameter is the scale factor, which has dimensions
of inverse velocity ($s/m$) and, following Eq.~(2)
of Ref.~\onlinecite{Viermann} is given by
\bea
a^2(t) &=& \sqrt{\frac{m^3}{8\pi \omega_z \hbar^3 \bar{n}_0^2}}
\frac{1}{\ab} \frac{\ab}{\as(t)},
\\
&=& 4.4\times 10^7 \frac{s^2}{m^2} \frac{\ab}{\as(t)},
\label{eq:asquaredred}
\eea
with $\as(t)$ the experimentally-controllable
s-wave scattering length, which we wrote as a ratio of
the Bohr radius $\ab = 5.29\times 10^{-11}{\rm m}$ 
for convenience.  In the second line we 
plugged in numbers from Ref.~\onlinecite{Viermann},
including
$\omega_z = 2\pi \times 1.6 {\rm kHz}$,
$\bar{n}_0 = 1.3\times 10^9/(cm^2)$, along with $m = 6.47\times 10^{-26}$kg the mass of a potassium-$39$ atom.

To get an estimate for $a_i$, the initial scale factor,
we use the fact that Viermann et al quote $400\ab$ as
the maximum scattering length used in their 
experiment.  Plugging this into Eq.~(\ref{eq:asquaredred})
gives $a_i = 331s/m$.  
To estimate the Hubble rate parameter $H$,  we search for conditions
that may increase the particle production.  To do this,  we note that 
the dependence of $|\beta|^2$ on $\ell$ comes via the eigenvalue
$h(\ell) = - \kappa \ell(\ell+1)$  
(see below Eq.~(\ref{eq:whyellm}))
with $\kappa = 4/R^2$ the curvature.  In fact, the 
dimensionless quantity appearing in the formula for $\beta$ is:
\be
\frac{\omega_k^i}{H}  = \frac{\sqrt{|h|}}{ a_i H } = \frac{240 {\rm s}^{-1}}{H}
\sqrt{\ell(\ell+1)/2},
\label{Eq:argument}
\ee
where to get the estimate on the right side we 
used the estimate $R = \sqrt{2}R_{TF}$ with 
$R_{TF}  = 25\mu m$ from Ref.~\onlinecite{Viermann}, which
they used for the hyperbolic spacetime curvature (but
we can expect to be reasonable here).  This formula tells
us that if $H$ is too small, then the argument Eq.~(\ref{Eq:argument})
will be large even for $\ell = 1$, leading to a small value of $|\beta|^2$.
To have a strong particle production, then, we pick $H$ so that 
this argument is not small; for concreteness we choose $H = 240{\rm s}^{-1}$. 
so the argument is unity at $\ell = 1$.

Our results for the particle production $|\beta|^2$ vs.
$\ell$, using these parameters and the expansion
time $t_f = 4m$s, are shown in 
Fig.~\ref{particleproduction}.  
We see that the particle production mostly decreases
with increasing $\ell$, with small oscillations, being
maximal for $\ell = 1$.  

Our next task is to study the entanglement associated with
this quantum particle 
production, a problem that is of interest in the present context of
analog systems~\cite{Robertson2017,Robertson20172,Chen:2021xhd}
but which also connects to the general question of how to probe
nontrivial many-body quantum correlations, a topic reviewed 
in Ref.~\onlinecite{Frerot}.
The possibility of observing entangled pair production in analog 
expanding BECs has also been investigated in a recent paper by
Agullo and collaborators~\cite{Agullo2024}, also focusing on the
Viermann et al setup.

\subsection{Two-mode squeezing and entanglement}
The particle production process described in the preceding corresponds to 
the production of pairs of particles with quantum numbers $m$, $\mb\equiv -m$.  In this section, we study the 
logarithmic negativity measure of entanglement, 
$E_{\mathcal{N}}$,  which is based on the Peres-Horodecki or PPT (Positivity of the Partial Transpose) 
criterion~\cite{Peres:1996dw,Plenio:2005cwa,Simon:1999lfr}.

Our calculation of $E_{\mathcal{N}}$ follows the book
of Serafini~\cite{Serafini} on entanglement in
Gaussian states (which applies since the Bogoliubov-type theory
used here amounts to a Gaussian approximation), 
as well as other recent works that
have computed the entanglement negativity in 
 Hawking radiation~\cite{Agullo:2021vwj,Brady:2022ffk}  and analog inflationary  systems~\cite{Bhardwaj2024}.

The first step is to write a matrix equation for
the relation Eq.~(\ref{eq:modoprel}) between  the final ($\hat{b}$) and initial ($\hat{a}$) mode 
operators: 
\bea
\begin{pmatrix}
  \bh_{m} \\  \bh_{m}^\dagger  \\ \bh_{\mb} \\ \bh_{\mb}^\dagger
\end{pmatrix}
= \begin{pmatrix} \alpha^* & 0 & 0 & \beta^*
  \\
  0 & \alpha & \beta & 0
  \\
  0& \beta^* & \alpha^* & 0
  \\
  \beta & 0 & 0 & \alpha
  \end{pmatrix}\begin{pmatrix}
  \ah_{m}\\  \ah_{m}^\dagger  \\ \ah_{\mb} \\ \ah_{\mb}^\dagger
\end{pmatrix},
\label{Eq:dynamics}
\eea
where we suppressed the subscript $k$ for simplicity.  Defining column vectors and
a matrix $\bS$, we can write the preceding as
$\Bh  = \bS \Ah$.  Here, $\bS$ satisfies $\bS \bJ \bS^T = \bJ$, with $\bJ = 
\begin{pmatrix} i\sigma_y  & \mathbb{0}
  \\
   \mathbb{0} &i\sigma_y
  \end{pmatrix}$
  the symplectic form.  Here, $\sigma_y$ is the standard Pauli matrix,
  so that $i\sigma_y = \begin{pmatrix} 0 & 1 \\ -1 & 0 \end{pmatrix}$.  

  Within the Bogoliubov
approximation used here, our system is Gaussian, which means that all information 
from the density matrix is contained within the covariance matrix~\cite{Serafini},
which, for the initial system, is written in the mode basis as:
\bea
&&\bsigma_A^{i} = \langle \Big( \Ah \Ah^T + (\Ah \Ah^T)^T\Big)\rangle , 
\\ &&=\!\!
\begin{pmatrix}
  \langle \{\ah_m, \ah_m\}\rangle &
 \langle \{\ah_m, \ah_m^\dagger\} \rangle &
  \langle \{\ah_m, \ah_{\mb}\} \rangle
  &
  \langle \{\ah_m, \ah_{\mb}^\dagger\} \rangle
  \\
  \langle \{\ah_m^\dagger, \ah_m\}\rangle
  & \langle \{\ah_m^\dagger, \ah_m^\dagger\} \rangle
 &
  \langle \{\ah_m^\dagger, \ah_{\mb}\} \rangle
  & \langle \{\ah_m^\dagger, \ah_{\mb}^\dagger\} \rangle
  \\
    \langle \{\ah_{\mb}, \ah_m\}\rangle 
  & \langle \{\ah_{\mb}, \ah_m^\dagger\} \rangle
&
  \langle \{\ah_{\mb}, \ah_{\mb}\} \rangle
  & \langle \{\ah_{\mb}, \ah_{\mb}^\dagger\} \rangle
  \\
  \langle \{\ah_{\mb}^\dagger, \ah_m\}\rangle
&
  \langle \{\ah_{\mb}^\dagger, \ah_m^\dagger\} \rangle
    & \langle \{\ah_{\mb}^\dagger, \ah_{\mb}\} \rangle
  & \langle \{\ah_{\mb}^\dagger, \ah_{\mb}^\dagger\} \rangle
\end{pmatrix}
,\nonumber 
\eea
with a similar form holding for the covariance matrix $\bsigma_B$  of the 
final mode operators. 

We now assume the initial system is in a thermal state at temperature $T$, i.e., a BEC with a Bose
distribution ($\nb= \frac{1}{{\rm e}^{\beta \hbar \omega_k^i}-1}$) of excited quasiparticles.  In that case, the initial covariance 
matrix is simply (here $\sigma_x = \begin{pmatrix} 0 & 1 \\ 1& 0\end{pmatrix}$):
\be
\label{eq:sigmaInMode}
\bsigma^{i}_A =   (1+2\nb)
\begin{pmatrix} \sigma_x & \mathbb{0}
  \\
\mathbb{0}  &  \sigma_x
  \end{pmatrix}.
\ee
The final covariance matrix after the particle-production is obtained
from $\bsigma_B^f = \bS \bsigma_A^i \bS^\dagger$:
\bea
&&\bsigma^{f}_B 
= (1+2\nb)\\
\nonumber 
&& \times 
\begin{pmatrix} 0 &|\alpha|^2+|\beta|^2
  & 2\alpha^*\beta^* &0
  \\
  |\alpha|^2+|\beta|^2& 0& 0& 2\alpha \beta
  \\
  2\alpha^*\beta^* & 0 & 0 &  |\alpha|^2+|\beta|^2
  \\
  0&  2\alpha \beta &   |\alpha|^2+|\beta|^2 & 0
  \end{pmatrix}
\eea
and possesses 
nontrivial off-diagonal structure due to the two-mode squeezing 
process.

To see how these correlations translate to entanglement, we change
from the mode basis to the $x$-$p$ basis by defining
\bse
\label{eq:xpmode}
\bea
\Xh_m &=& \frac{1}{\sqrt{2}} \Big(\ah_m + \ah_m^\dagger\Big),
\\
\Ph_m &=& -\frac{i}{\sqrt{2}}\Big(\ah_m - \ah_m^\dagger\Big),
\label{Eq:peedef}
\eea
\ese
in the initial system (and similarly for the final system,
with $\bh$ mode operators on the right).
This further motivates defining the $\Xh-\Ph$ covariance matrix
\bea
\label{Eq:covarianceDefinition}
&&\bsigma  = 
\\
&&
\begin{pmatrix}
  \langle \{\Xh_m, \Xh_m\} \rangle & \langle \{\Xh_m, \Ph_m\} \rangle &
  \langle \{\Xh_m, \Xh_{\mb}\} \rangle & \langle \{\Xh_m ,\Ph_{\mb}\} \rangle
  \\
  \langle \{\Ph_m, \Xh_m\} \rangle & \langle \{\Ph_m, \Ph_m\} \rangle
  &\langle \{\Ph_m, \Xh_{\mb}\} \rangle & \langle \{\Ph_m, \Ph_{\mb}\} \rangle
   \\
   \langle \{\Xh_{\mb}, \Xh_m\} \rangle & \langle \{\Xh_{\mb}, \Ph_m\} \rangle &
   \langle \{\Xh_{\mb}, \Xh_{\mb}\} \rangle & \langle \{\Xh_{\mb}, \Ph_{\mb}\} \rangle
   \\ \langle \{\Ph_{\mb}, \Xh_m\} \rangle & \langle \{\Ph_{\mb}, \Ph_m\} \rangle
   &\langle \{\Ph_{\mb}, \Xh_{\mb}\} \rangle & \langle \{\Ph_{\mb}, \Ph_{\mb}\} \rangle
  \end{pmatrix}.
  \nonumber 
\eea
In the initial case, we obtain the initial covariance matrix from $\bsigma^{i}_A$ in 
Eq.~(\ref{eq:sigmaInMode}) using 
$\bsigma^{i} = \Uh \bsigma_A^i \Uh^\dagger$ with the unitary matrix 
$\Uh = \bone \otimes  \frac{1}{\sqrt{2}}\begin{pmatrix} 1 & 1 \\
-i & i \end{pmatrix}$ following from Eqs.~(\ref{eq:xpmode}).  Similarly, we have
$\bsigma^f =  \Uh \bsigma_B^f \Uh^\dagger$ for the final covariance matrix
in the mode basis.

Since they describe quantum systems, the initial
and final covariance matrices are constrained
by the uncertainty principle, which implies that
the symplectic eigenvalues $\nu_n$ (obtained
from the  eigenvalues of $i\bJ \bsigma$)
must satisfy $\nu_n\geq 1$, with $\bJ$ the 
symplectic form given above. A direct calculation
indeed shows this inequality indeed holds for the symplectic eigenvalues 
$\bsigma^i$ and $\bsigma^f$.

To access entanglement, we examine this inequality
after performing a partial transpose operation on 
one of the two modes in our two-mode system.
Applying this operation to mode $\mb$ for concreteness, 
the
partial transpose on system $\mb$ takes the form 
$\ah_{\mb} \leftrightarrow \ah^\dagger_{\mb}$, or,
in the $\Xh-\Ph$ basis, simply mapping 
$\Xh_{\mb} \to \Xh_{\mb}$ and 
$\Ph_{\mb} \to - \Ph_{\mb}$ as seen from 
Eqs.~(\ref{eq:xpmode}).

In terms of the covariance
matrix, this operation amounts to 
the mapping $\bsigma_{\rm PT} = \Th \bsigma \Th$ 
with $\Th = {\rm diag.}(1,1,1,-1)$.  Nonseparability of
the system is signaled by a violation of the above
inequality after the partial transpose operation, i.e., 
one of the symplectic eigenvalues crosses unity.  
For the two-mode system considered here, there are two symplectic
eigenvalues $(1+2\nb)(\sqrt{1+|\beta|^2}\pm |\beta|)^2$, and only
the $-$ case can be less than unity.  This gives, for the 
logarithmic negativity measure of entanglement (reintroducing 
the dependence on the quantum number $\ell$):
\bea
\label{LogNeg}
&&
E_{\mathcal{N}}[\ell]
\\
&&= \text{Max}\Big(0,-\log_{2}\Big[(1+2n_{\text{B}})\left(\sqrt{1+|\beta_{\ell}|^2}-|\beta_{\ell}|)^{2}\right)\Big],
\nonumber 
\eea
\begin{figure}
    \centering
    \includegraphics[width=\linewidth]{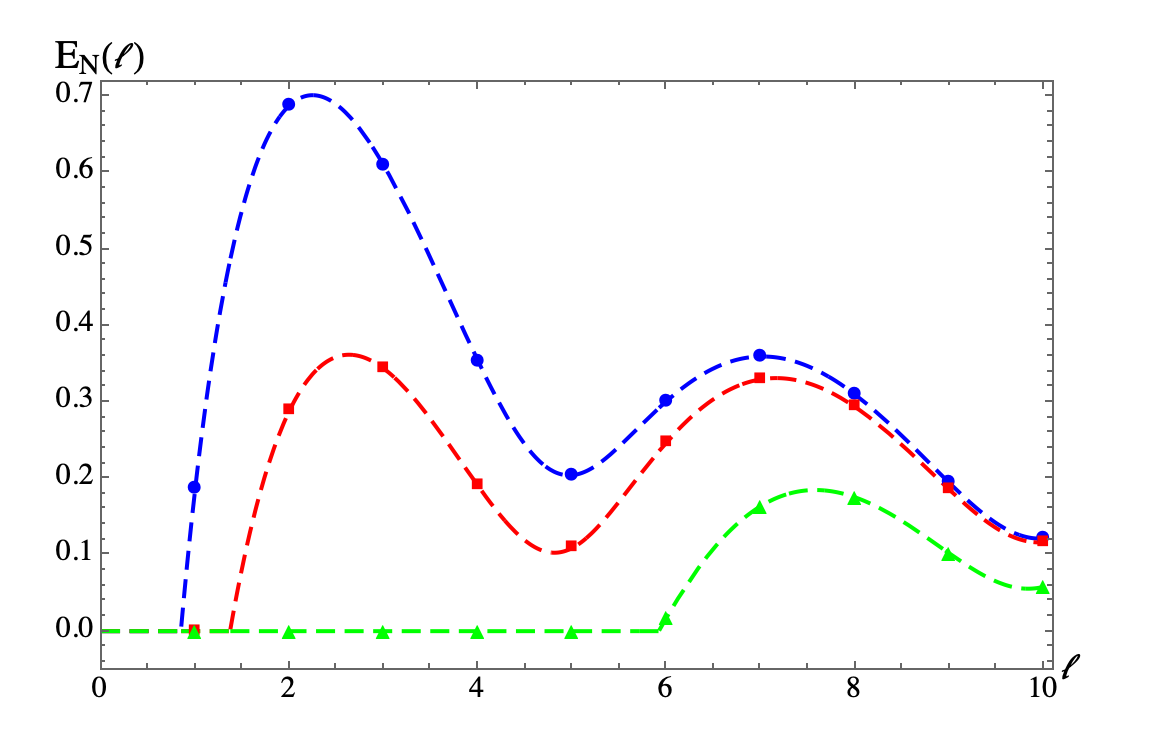}
    \caption{Logarithmic negativity \(E_N [\ell]\) vs the mode index \(\ell\), for various values of the environment temperature, \(T = 2nK\) (blue circles), \(T = 3nK\) (red squares), and \(T = 5nK\)(green triangles). This plot was generated using the same \(a_0\) and \(H\) as in Fig. \ref{particleproduction}, with the expansion time being \(4ms\). We can see that the entanglement of modes at low energies is suppressed by environment temperature.}
    \label{logneg}
\end{figure}
In Fig.~\ref{logneg} we show our result for $E_{\mathcal{N}}[\ell]$ 
for various values of the environmental temperature.  We see that, at
sufficiently low temperature, the Viermann et al setup should produce
quantum entangled pairs of particles with quantum numbers 
$\ell,m$ and $\ell,\bar{m}$.  We find that increasing temperature
suppresses entanglement for low-$\ell$ particles.  

One open question is how the entanglement of such pairs could be 
directly probed experimentally.  Since $E_{\mathcal{N}}[\ell]$ is
related to the covariance matrix that involves correlation functions,
one way to do this is to experimentally measure all such correlation functions and hence work out the entire covariance matrix, a possibility
studied in Ref.~\onlinecite{Bhardwaj2024} for the case of an 
expanding toroidal BEC.  However, a less experimentally-intensive 
probe would be desirable,
a question we leave for future work.  

\section{Conclusion and Perspectives}
\label{sec:concl}
In this work, we investigated classical and quantum waves in a spatially
inhomogeneous two-dimensional Bose-Einstein condensate BEC engineered to realize a spherical FLRW geometry.  
 Our work was inspired by the recent work by Viermann et al.~\cite{Viermann}, who manufactured disk-shaped BECs in to realize both spherical and hyperbolic 2+1 dimensional FLRW universes (although we focused on the
 spherical case here).

 While the experiment of Viermann et al considered a polynomial-in-time ramp in the scale factor, and observed enhanced fluctuations as indications for particle production, our focus was on an exponential expansion.  In the case of
 classical wave propagation, we found that the scale factor discontinuity
 at the start and end of expansion created additional cusp-like waves 
 traveling opposite to the main wave motion.  We also found that an
 initial vacuum state would yield quantum particle production -- or 
 amplification of the quantum vacuum~\cite{Nation:2011dka}. 
 We made an estimate of the predicted number of particles created, using 
 parameters based on the experiments of Ref.~\onlinecite{Viermann}.

We close with some ideas for future directions for investigation. 
While our particle production calculations focused on the case of an initial
equilibrium BEC, it would be interesting to study other initial states (such as 
coherent states or squeezed initial states).  
Since
the Viermann experiments also realized a hyperbolic FLRW universe, it would
be natural to extend our calculations to that case.  One qualitative
difference is that the dispersion is no longer gapless.  
We also note that the mathematical physics literature has investigated 
the Greens functions for hyperbolic and curved spacetimes (See, e.g., 
Ref.~\onlinecite{Durand2023}) and it would be interesting to connect 
our numerical approach to these calculations and explore further
properties of wave propagation in emergent curved-spacetime 
geometries.

\section{Acknowledgments}
JAC acknowledges support  from the National Science Foundation from its REU Site in Physics and Astronomy (NSF Grant No. 2150445) at Louisiana State University.
RZ and DES acknowledge support from the National Science Foundation under Grant PHY-2208036.

\appendix

\section{Equations left from main text}
\label{sec:appendixA} 
In this appendix, we present our explicit formulas for various
functions described in the main text.  We start with
$A_i$ and $B_i$ appearing in
Eq.~(\ref{eq:peetoo}) for the function $p^{II}_k(t)$
in the wave equation for an exponential expansion:
\begin{equation}
\begin{gathered}
A_i = - \frac{\left(\sin{\left(t_{i} w_{k}^{i} \right)} Y_{0}\left(\frac{w_{k}^{i}}{H}\right) + \cos{\left(t_{i} w_{k}^{i} \right)} Y_{1}\left(\frac{w_{k}^{i}}{H}\right)\right) e^{H t_{i}}}{a_{i} \sqrt{h} \left(J_{0}\left(\frac{w_{k}^{i}}{H}\right) Y_{1}\left(\frac{w_{k}^{i}}{H}\right) - J_{1}\left(\frac{w_{k}^{i}}{H}\right) Y_{0}\left(\frac{w_{k}^{i}}{H}\right)\right)},\\
B_i = \frac{\left(\sin{\left(t_{i} w_{k}^{i} \right)} J_{0}\left(\frac{w_{k}^{i}}{H}\right) + \cos{\left(t_{i} w_{k}^{i} \right)} J_{1}\left(\frac{w_{k}^{i}}{H}\right)\right) e^{H t_{i}}}{a_{i} \sqrt{h} \left(J_{0}\left(\frac{w_{k}^{i}}{H}\right) Y_{1}\left(\frac{w_{k}^{i}}{H}\right) - J_{1}\left(\frac{w_{k}^{i}}{H}\right) Y_{0}\left(\frac{w_{k}^{i}}{H}\right)\right)},
\end{gathered}
\end{equation}
Next we present 
the coefficients $A_f$ and $B_f$ in the
wave equation solution (see Eq.~(\ref{Eq:peeEff})
after the end of expansion:
\begin{multline}
A_f = - a_{f} \omega_{k}^{I}  e^{- 2 H t_{f}}\Biggl[A_{i} a_{f} e^{H t_{i}} \cos{\left( \omega_{k}^{f} t_f \right)} J_{0}\left(\frac{\omega_{k}^{I} e^{- H \left(t_{f} - t_{i}\right)}}{H}\right) \\- A_{i} a_{i} e^{H t_{f}} \sin{\left( \omega_{k}^{f} t_f \right)} J_{1}\left(\frac{\omega_{k}^{I} e^{- H \left(t_{f} - t_{i}\right)}}{H}\right) +  B_{i} a_{f} e^{H t_{i}} \cos{\left( \omega_{k}^{f} t_f \right)} \\Y_{0}\left(\frac{\omega_{k}^{I} e^{- H \left(t_{f} - t_{i}\right)}}{H}\right) - B_{i} a_{i} e^{H t_{f}} \sin\left( \omega_{k}^{f} t_f \right) Y_{1}\left(\frac{\omega_{k}^{I} e^{- H \left(t_{f} - t_{i}\right)}}{H}\right)  \Biggr],
\end{multline}
\begin{multline}
B_f =  a_{f} \omega_{k}^{I}e^{- 2 H t_{f}} \Biggl[A_{i} a_{f} e^{H t_{i}} \sin{\left( \omega_{k}^{f} t_f \right)} J_{0}\left(\frac{\omega_{k}^{I} e^{- H \left(t_{f} - t_{i}\right)}}{H}\right) \\+ A_{i} a_{i} e^{H t_{f}} \cos{\left( \omega_{k}^{f} t_f \right)} J_{1}\left(\frac{\omega_{k}^{I} e^{- H \left(t_{f} - t_{i}\right)}}{H}\right) +  B_{i} a_{f} e^{H t_{i}} \sin{\left( \omega_{k}^{f} t_f \right)} \\Y_{0}\left(\frac{\omega_{k}^{I} e^{- H \left(t_{f} - t_{i}\right)}}{H}\right) + B_{i} a_{i} e^{H t_{f}} \cos{\left( \omega_{k}^{f} t_f \right)} Y_{1}\left(\frac{\omega_{k}^{I} e^{- H \left(t_{f} - t_{i}\right)}}{H}\right)\Biggr].
\end{multline}
We also present our formula for the quantum particle
production parameter $\beta$:
\begin{multline}
\beta= \left(-i J_{0} \left( \frac{e^{-H t_f} \omega_k^i}{H} \right) + J_{1} \left( \frac{e^{-H t_f} \omega_k^i}{H} \right) \right) \\
\cdot \left( Y_{0} \left( \frac{\omega_k^i}{H} \right) + i Y_{1} \left( \frac{\omega_k^i}{H} \right) \right) 
+ i \left( J_{0} \left( \frac{\omega_k^i}{H} \right) + i J_{1} \left( \frac{\omega_k^i}{H} \right) \right) \\
\cdot \left. \left( Y_{0} \left( \frac{e^{-H t_f} \omega_k^i}{H} \right) + i Y_{1} \left( \frac{e^{-H t_f} \omega_k^i}{H} \right) \right) \right).
\label{eq:afourforbeta}
\end{multline}


\end{document}